\documentclass[preprint, 
amsmath, 
amssymb, 
aps]{revtex4-2}

\usepackage{color}
\usepackage{graphicx}
\usepackage{dcolumn}
\usepackage{diagbox}
\usepackage[mathlines]{lineno}
\usepackage{booktabs}
\usepackage{adjustbox}\usepackage{mathtools}
\usepackage{array}
\usepackage{makecell,booktabs}
\usepackage{stackengine}
\setstackEOL{\cr}
\usepackage{comment}

\newcommand\nablab{\boldsymbol{\nabla}}
\newcommand\dd{\mathrm{d}}
\newcommand\DD{\mathrm{D}}
\newcommand\ii{\mathrm{i}}
\newcommand\uvect[1]{\widehat{\mathbf{#1}}}

\begin{document}

\preprint{APS/123-QED}

\title{
Validity of sound-proof approximations for magnetic buoyancy}

\author{J. B. Moss}
 \email{j.moss2@newcastle.ac.uk}
\author{T. S. Wood}
\author{P. J. Bushby}
\affiliation{Newcastle University, Newcastle-upon-Tyne, NE1 7RU, UK}

\date{\today}

\begin{abstract}
The presence of acoustic waves in models of compressible flows can present complications for analytical and numerical analysis.
Therefore, several methods have been developed to filter out these waves,
leading to various ``sound-proof'' models,
including the Boussinesq, anelastic and pseudo-incompressible models.
We assess the validity of each of these approximate models for describing magnetic buoyancy in the context of the solar interior.
A general sound-proof model is introduced and compared to the fully compressible system in a number of asymptotic regimes, including both non-rotating and rotating cases.
We obtain specific constraints
that
must be satisfied in order that the model captures the leading-order behaviour of the fully compressible system.
We then discuss which of the existing sound-proof models satisfy these constraints, and in what parameter regimes.
We also present a variational derivation of the pseudo-incompressible MHD model, demonstrating its underlying Hamiltonian structure.
\end{abstract}

\maketitle

\section{Introduction \label{sec: introduction}}
    To model buoyancy-driven flows, in many astrophysical and geophysical contexts it is essential to include effects of compressibility and stratification.
    However, in many cases the dynamics of interest occur on a timescale that is much longer than the acoustic timescale,
    i.e., the time taken for a sound wave to traverse the fluid. 
    Most models, whether theoretical or numerical, therefore make some kind of sound-proof approximation (e.g., Boussinesq or anelastic)
    in which sound waves are filtered out of the governing equations.
    In addition to simplifying the mathematics, such approximations often allow much larger time-steps to be taken in numerical calculations (with the short acoustic timescale no longer providing a constraint), whilst still accurately resolving the important dynamics.
    However, each of these approximations is derived under certain assumptions that may not hold in the system of interest.
    In particular, the Boussinesq approximation is derived under the assumption of small length-scales,
    and the anelastic approximation is derived assuming small perturbations
    to a state with uniform entropy.
    
    Magnetic buoyancy --- the tendency for regions of strong magnetic flux to be less dense than their surroundings ---
    potentially provides a stringent test problem for any sound-proof model,
    because it involves significant perturbations to the fluid pressure and density,
    on length-scales that are typically long in the direction of the magnetic field but short in the other directions.
    Yet there has been relatively little work done to compare how accurately different sound-proof models describe the magnetic buoyancy instability, outside of specific asymptotic regimes.
    The main scientific interest in the magnetic buoyancy instability comes from the solar interior,
    where buoyant magnetic structures rise through the Sun's convective envelope and emerge at the surface.
    These buoyant structures are believed to originate in the tachocline below the convection zone,
    where strong magnetic fields are generated by differential rotation.
    In the lower part of the tachocline the temperature gradient
    is strongly subadiabatic,
    inhibiting the magnetic buoyancy instability until the field becomes sufficiently strong.
    Most global numerical studies of the solar interior have employed the anelastic approximation.
    However, it is unclear
    how accurately the anelastic approximation captures the onset of instability in the presence of such large entropy variations.
    Indeed, under these conditions the meaning of the anelastic approximation becomes somewhat ambiguous,
    because there are different formulations that only become equivalent in the asymptotic limit of adiabatic stratification
    \citep[e.g.,][]{Berkoff10}.
    
    Our goal in this paper is to determine which sound-proof models accurately describe the magnetic buoyancy instability,
    and in precisely which parameter regimes.
    Our approach is similar to that of \citet{Berkoff10},
    who numerically solved the linearised equations
    describing perturbations to specific background states,
    and compared the growth rates they obtained using the anelastic and fully compressible models.
    However, here we consider several asymptotic parameter regimes
    in order to obtain analytical solutions for more general background states.
    This allows us to identify more precisely the conditions under which different sound-proof approximations reproduce the fully compressible results.
    Note that we will only be considering sound-\textit{proof} models --- that is, models in which sound waves are removed from the governing equations --- rather than alternative models that slow down sound waves \citep{iijima19}
    or damp them using implicit timestepping \citep{Goffrey-etal17}.
    
    Whilst the relative merits of the different sound-proof approximations have been extensively studied in the context of various hydrodynamic problems \cite{ 
    Bannon95, 
    Lilly96,
    Davies03,
    Brown12,
    WoodBushby16}, 
    there are still a number of important open questions regarding the applicability of such approximations in the context of magnetohydrodynamics.
    It is one of these open questions (i.e.\ the extent to which such approximations can be used to describe magnetic buoyancy) that provides the motivation for this paper.
    Recent work by \citet{Wilczynski22} presents a similar analysis focused on the analestic model,
    and restricted to nearly-adiabatic parameter regimes, for which the anelastic model is asymptotically valid.
    Here, we consider a much wider class of sound-proof models, and we also consider a wider range of parameter regimes.

In Section \ref{sec: FC intro} we define the geometry and essential notation of the problem,
and present the linearised equations that describe magnetic buoyancy.
Section \ref{sec: existing sound-proof} briefly reviews the best-known sound-proof models, namely the magneto-Boussinesq, anelastic, and pseudo-incompressible models,
before Section \ref{sec: sound-proof} introduces a general linearised sound-proof model
that includes several adjustable coefficients.
This general model includes each of the sound-proof models just mentioned as special cases.
In Section \ref{sec: non-rotating} we compare the linear stability properties of our sound-proof model with the fully compressible system in the absence of rotation;
Section~\ref{sec: rotating} presents a similar analysis for the rotating fluid.
In each case we identify the leading-order dynamics in both the fully compressible system and our general sound-proof model,
both with and without thermal relaxation.
Comparison between these two systems yields constraints on the coefficients in our model which in turn tell us which models are 
applicable
in which regimes.
A summary and discussion of these results is presented in Section~\ref{sec: conclusion}.

\section{Fully Compressible Equations} \label{sec: FC intro}

Before discussing sound-proof models in the next section,
we first introduce the fully compressible equations of magnetohydrodynamics (MHD).
Under the local Cartesian ($f$-plane) approximation, with uniform angular velocity $\mathbf{\Omega}$, these are
\begin{align}
    \rho \frac{\DD \mathbf{u}}{\DD t} + 2 \rho \, \mathbf{\Omega} \times \mathbf{u} &= - \nablab p - g \rho \, \uvect{z} + \frac{1}{\mu_0} (\nablab \times \mathbf{B}) \times \mathbf{B}\,,
    \label{eq:full_momentum} \\
    \frac{\partial \mathbf{B}}{\partial t} &= \nablab\times(\mathbf{u} \times \mathbf{B})\,,
    \label{eq:full_induction} \\
    \frac{\DD \rho}{\DD t} + \rho \nablab \cdot \mathbf{u} &= 0\,,
    \label{eq:full_continuity} \\
    \rho T\frac{\DD s}{\DD t} &= Q\,.
    \label{eq:full_heat}
\end{align}
Here, $Q$ represents the sum of all diabatic heating processes, such as thermal diffusion, $\tfrac{\DD}{\DD t} \equiv \tfrac{\partial }{\partial t} + \mathbf{u} \cdot \nablab$ is the material derivative, $\rho$ is the fluid density, $\mathbf{u} = (u_x, u_y, u_z)$ is the fluid velocity, $p$ is the pressure, $\mu_0$ is the magnetic permeability, $\mathbf{B}$ is the magnetic field, $s$ is the specific entropy, $T$ is the temperature, and $-g\uvect{z}$ is the gravitational acceleration, assumed to be uniform.
The equations are closed using an equation of state that allows us to express,
for example, $\rho$ and $T$ in terms of $p$ and $s$.
Note that the Lorentz force can be decomposed into the sum of magnetic pressure and magnetic tension terms as $\frac{1}{\mu_0} (\nablab \times \mathbf{B}) \times \mathbf{B} = - \frac{1}{2\mu_0} \nablab |\mathbf{B}|^2 + \frac{1}{\mu_0} (\mathbf{B} \cdot \nablab )\mathbf{B}$. We can then define the total pressure, $\pi$, to be the sum of the gas pressure and the magnetic pressure, i.e., $\pi \equiv p + \frac{1}{2 \mu_0} |\mathbf{B}|^2$.

For simplicity we have neglected viscosity and magnetic diffusion;
in general these are expected to play only a minor role in magnetic buoyancy under solar conditions
(although they can result in double-diffusive-type instabilities, e.g., \citep{HughesWeiss95}).
Under astrophysical conditions, the dominant diabatic process is usually thermal diffusion, so that in equation~(\ref{eq:full_heat}) we have
$Q = \nablab \cdot (\rho c_p\kappa \nablab T)$, where $c_p$ is the specific heat capacity, $\kappa$ is the thermal diffusivity, and $T$ is the temperature.
However, in what follows we will instead introduce Newtonian cooling, with $Q = \rho c_p\alpha(T_0 - T)$,
where $T_0(z)$ is the equilibrium temperature profile
and $\alpha$ is the cooling rate.
This simplifies the analysis by reducing the number of vertical derivatives in the linearised equations.
In what follows, we are mostly concerned with the limits $\alpha \to 0$
and $\alpha \to \infty$,
the latter of which reproduces the ``fast thermal relaxation'' regime considered by \citet{Gilman70}, in which temperature perturbations are vanishingly small.
In both of these limits, the exact mechanism of thermal relaxation becomes irrelevant.

Within the tachocline, beneath the solar convection zone,
there is strong rotational shear that is expected to generate strong toroidal magnetic field.
Into our local Cartesian model we will therefore introduce a layer of magnetic field oriented in the $x$-direction, representing azimuth,
whose strength varies with altitude, $z$.
We will consider linear perturbations to a background state in
magneto-hydrostatic balance
in order to determine the conditions under which this magnetic layer becomes buoyantly unstable.
From previous studies \citep[e.g.,][]{Gilman70,Acheson79,Hughes07article} we expect the growth rate of the instability to be of the order of weeks or longer,
whereas the acoustic timescale in the solar interior is of the order of minutes.
Moreover, the Alfv\'en speed, $v \equiv |\mathbf{B}|/\sqrt{\mu_0\rho}$,
is expected to be only a very small fraction of the sound speed, $c$, given by $c^2 \equiv \left( \frac{\partial p}{\partial \rho} \right)_s$,
which suggests that the instability ought to be captured to high accuracy
by some form of sound-proof model.
This suggestion forms the motivation for the present study.

\subsection{Linearised magnetic buoyancy equations \label{sec: fully compressible} }

We begin by taking the fully compressible equations
(\ref{eq:full_momentum})--(\ref{eq:full_heat})
and considering linear perturbations to a static background state containing a horizontal layer of magnetic field oriented in the $x$-direction,
$\mathbf{B} = B_0(z)\uvect{x}$.
Each quantity is expanded as $f = f_0(z) + f_1(x,y,z,t)$,
where $f_0$ is the background value and $f_1$ is the linear perturbation.
The background state is in magneto-hydrostatic balance, i.e.,
\begin{equation}
  \frac{\dd\pi_0}{\dd z} =
  \frac{\dd}{\dd z}\left(p_0 + \frac{B_0^2}{2\mu_0}\right) = - g\rho_0\,.
\end{equation}
This implies a relation between the scale heights of density, entropy and magnetic field:
\begin{equation}
  g = c^2 H_\rho^{-1} - c^2H_s^{-1} + v^2H_B^{-1}\,,
  \label{eq:g}
\end{equation}
where (as before) $c$ is the sound speed and $v$ is the Alfv\'en speed, both defined in terms of the background state.
For a general equation of state, the scale heights in equation~(\ref{eq:g}) are defined as:
\begin{align*}
    H_\rho^{-1} &\equiv - \frac{\dd}{\dd z} \ln \rho_0\,, \\
    H_s^{-1} &\equiv - \left( \frac{\partial \ln \rho}{\partial s} \right)_p \frac{\dd s_0}{\dd z}\,, \\
    H_B^{-1} &\equiv - \frac{\dd}{\dd z} \ln B_0\,,
\end{align*}
where the partial derivative is evaluated in the background state.
Note that, in many studies of magnetic buoyancy, the Brunt-V\"ais\"al\"a frequency, $N$, is defined as
\begin{equation}
  N^2 = - g\left( \frac{\partial \ln \rho}{\partial s} \right)_p \frac{\dd s_0}{\dd z} = g H_s^{-1}\,.
\end{equation}
However, the physical meaning of $N$ becomes ambiguous in the presence of a magnetic field, so we prefer to work with $H_s$ instead.

When comparing results from the fully compressible equations
with results from any sound-proof model,
care must be taken in the interpretation of the background state.
In particular, most sound-proof models first introduce a non-magnetic
background state in hydrostatic balance, and regard the entire magnetic field as a perturbation to this.
For consistency, we will therefore only consider background states that are
weakly magnetised, in the sense that the magnetic pressure force is negligible in comparison with gravity.
This means that the final term in equation~(\ref{eq:g}) can be neglected, at least in a leading-order analysis.

The linearised fully compressible equations are
\begin{align}
    \rho_0 \frac{\partial}{\partial t}\mathbf{u} + 2 \rho_0 \mathbf{\Omega} \times \mathbf{u} &= - g \rho_1 \uvect{z} - \nablab \pi_1 + \frac{1}{\mu_0}\mathbf{B}_0\cdot\nablab\mathbf{B}_1 + \frac{1}{\mu_0} \mathbf{B}_1\cdot\nablab\mathbf{B}_0\,,
    \label{eq:momentum} \\
    \frac{\partial}{\partial t}\mathbf{B}_1 &= \nablab \times (\mathbf{u} \times \mathbf{B}_0)\,, \\
    \frac{\partial}{\partial t}\rho_1 &= - \nablab \cdot (\rho_0 \mathbf{u})\,,
    \label{eq:continuity} \\
    \left( \frac{\partial \rho}{\partial s} \right)_p \frac{\partial}{\partial t}s_1 - \rho_0 H_s^{-1} u_z
    &=
    \left(\frac{\partial\rho}{\partial T}\right)_p \frac{Q_1}{\rho_0 c_p} , \label{eq: energy}\\
    \pi_1 &= p_1 + \frac{1}{\mu_0} \mathbf{B}_0\cdot\mathbf{B}_1\,, \\
    \left( \frac{\partial \rho}{\partial T} \right)_p T_1 &= \left( \frac{\partial \rho}{\partial s} \right)_p s_1 - \frac{\gamma - 1}{c^2} p_1\,, \label{eq:temp}\\
    \rho_1 &= \left( \frac{\partial \rho}{\partial s} \right)_p s_1 + \frac{1}{c^2} p_1\,,
    \label{eq:EoS}
\end{align}
where 
$\gamma$ is the adiabatic index
and $\pi_1$ is the perturbation to the total (i.e., gas plus magnetic) pressure.
In equation (\ref{eq: energy}), $Q_1$ represents perturbations to the diabatic heating;
in the case of Newtonian cooling, we have $Q_1 = - \rho_0 c_p\alpha T_1$.
Note that these equations are valid for an arbitrary equation of state.
The form of equations~(\ref{eq:temp})--(\ref{eq:EoS})
has been chosen so that the temperature and entropy perturbations,
$T_1$ and $s_1$,
have the same prefactors as in equation (\ref{eq: energy}),
which simplifies the subsequent algebra.

\section{Existing sound-proof models \label{sec: existing sound-proof} }

We now briefly review the most commonly-used sound-proof models for astrophysical applications:
the Boussinesq, anelastic, and pseudo-incompressible models.
(We do not consider the family of quasi-hydrostatic models often used to model shallow atmospheres \citep[e.g.,][]{MillerWhite84,ArakawaKonor09,DubosVoitus14},
because these are not suited to describing buoyancy processes
with a small horizontal scale,
which are the motivation for the present study.)
More details on the derivation of each of these models can be found in Appendix~\ref{sec: model-derivation}.
Here we will present each of these models in the form that they are usually found in the literature;
in Section~\ref{sec: sound-proof} we will introduce a very general, linearised sound-proof model for which each of these models arises as a special or limiting case.

\subsection{Boussinesq}
The Boussinesq model was first introduced by \citet{Oberbeck1879}
and \citet{Boussinesq03}, and later formalised by \citet{Jeffreys30}. 
The approximation amounts to assuming that,
on small length-scales, pressure variations do not produce significant changes in density.
This allows pressure perturbations to be neglected in the relations between other thermodynamic variables,
and density variations are neglected everywhere except in the buoyancy force
\cite{SpiegelVeronis60,Mihaljan62,GrayGiorgini76}.
As a result, the pressure perturbation effectively acts as a Lagrange multiplier that enforces a zero-divergence constraint on the fluid velocity, i.e.,~incompressibility.
The simplest way to extend the Boussinesq model to include a magnetic field is simply to add the Lorentz force to the momentum equation, and simultaneously solve the magnetic induction equation in its usual MHD form, i.e., equation~(\ref{eq:full_induction}).
However, this alone does not capture the effects of magnetic buoyancy.
\citet{SpiegelWeiss82} extended the Boussinesq model to include magnetic buoyancy by
(a)~assuming small variations to the \emph{total} (i.e.,~gas plus magnetic) pressure,
and (b)~allowing for density variations in the induction equation.
This magneto-Boussinesq approximation was later rederived more rigorously by \citet{Corfield84},
using formal asymptotics. 
The resulting equations, omitting all nonlinear terms for simplicity, can then be written as 
\begin{align}
    \rho_0\frac{\partial}{\partial t}\mathbf{u}
    + 2 \rho_0 \mathbf{\Omega} \times \mathbf{u} &= - g \rho_1 \uvect{z} - \nablab_\perp \pi_1 + \frac{1}{\mu_0}\mathbf{B}_0\cdot\nablab\mathbf{B}_1 - \frac{1}{\mu_0}H_B^{-1}B_{1z}\mathbf{B}_0\,,
    \label{eq:momentum_mb} \\
    \frac{\partial}{\partial t}\mathbf{B}_1 &=
    \mathbf{B}_0 \cdot \nablab \mathbf{u} + (H_B^{-1} - H_\rho^{-1})u_z \mathbf{B}_0 \,, \\
    \nablab_\perp \cdot \mathbf{u} &= 0\,,
    \label{eq:continuity_mb} \\
    \frac{\partial}{\partial t}s_1
    + \mathbf{u}\cdot\nablab s_0
    &=
    \frac{Q_1}{\rho_0 T_0} , \label{eq: energy_mb}\\
    p_1 + \frac{1}{\mu_0} \mathbf{B}_0\cdot\mathbf{B}_1 &= 0 \,, 
\end{align}
where $\rho_0$, $T_0$ and $\mathbf{B}_0$ are taken to be constant, and where the subscript $\perp$ represents components that are perpendicular to $\mathbf{B}_0$.
The thermodynamic relations~(\ref{eq:temp})--(\ref{eq:EoS}) are retained, except that the coefficients are usually taken to be constant.
The derivation assumes small length-scales in the directions perpendicular to the magnetic field (see Appendix~\ref{sec: model-derivation} for details),
and as a result the usual solenoidality condition for the field is replaced by
$\nablab_\perp\cdot\mathbf{B}_1 = 0$.
The fact that the orientation of the magnetic field must be known \textit{a priori}
makes the magneto-Boussinesq model unsuitable for numerical simulations with complex field geometries.

\subsection{Anelastic}

The anelastic model was first introduced by \citet{OguraPhillips62},
extending earlier work by \citet{Batchelor54}.
They showed that the small length-scale assumption used in the Boussinesq model can be dropped provided that the
fluid has almost uniform specific entropy.
Their model formally assumes small (though nonlinear) perturbations
to a background state with exactly uniform entropy,
i.e., adiabatic stratification.
Such a state is rarely achieved in any physical system, however,
and many subsequent works have therefore adapted the anelastic model to allow the specific entropy to be a slowly varying function of altitude
\citep[e.g.][]{Gough69,GilmanGlatzmaier81,LippsHemler82}.
These different formulations are equivalent only in the asymptotic limit of adiabatic stratification.
Today, the term ``anelastic'' is generally used to refer to any fluid model in which the continuity equation~(\ref{eq:full_continuity})
is replaced with a velocity constraint
\begin{equation}
  \nablab\cdot(\rho_0\mathbf{u}) = 0,
\end{equation}
where $\rho_0(z)$ represents the density of the background state
\citep{BraginskyRoberts07}.

Here, we will consider two particular formulations of the anelastic model.
Following \citet{WoodBushby16},
we refer to these as GGG (after \citet{Gough69,GilmanGlatzmaier81})
and LBR (after \citet{Lantz92,BraginskyRoberts95}).
The LBR model makes an additional approximation by replacing temperature diffusion with entropy diffusion;
this approximation can be justified by several different arguments
(see Appendix~\ref{sec: model-derivation})
but is not formally (i.e., asymptotically) valid.
Nonetheless, it has the benefit that the LBR equations can be solved without explicitly calculating either pressure or temperature,
and for this reason the LBR model has become the standard implementation in astrophysical applications \citep{Jones-etal11}.

The anelastic model is easily extended to include a magnetic field; provided that $v \ll c$, the only change is that the Lorentz force now appears in the momentum equation \citep{Glatzmaier84}.
The induction equation retains its flux-conservative form (\ref{eq:full_induction}), so $\nablab \cdot \mathbf{B}$ remains exactly zero.

The nonlinear GGG anelastic equations can be written as  
\begin{align}
    \rho_0 \frac{\DD \mathbf{u}}{\DD t} + 2 \rho_0 \mathbf{\Omega} \times \mathbf{u} &= - g \rho_1 \uvect{z} - \nablab p_1 + \frac{1}{\mu_0} ( \nablab \times \mathbf{B} ) \times \mathbf{B}\,, \label{eq: GGG_momentum}\\
    \nablab \cdot (\rho_0 \mathbf{u}) &= 0 \,, \label{eq: GGG_continuity}\\
    T_0\frac{\DD s_1}{\DD t} + (T_0 + T_1)\mathbf{u}\cdot\nablab s_0 &= \frac{Q}{\rho_0} \label{eq: GGG_energy} \,, 
\end{align}
which are solved alongside the induction equation~(\ref{eq:full_induction})
and the linearised equation of state~(\ref{eq:temp})--(\ref{eq:EoS}).
The left-hand side of equation~(\ref{eq: GGG_energy}) includes a term proportional to the temperature perturbation, $T_1$,
which is formally negligible under the assumptions of the anelastic derivation.
This term was retained in the model of \citet{Gough69}, in order to achieve a form of energy conservation,
but was neglected in the model of \citet{GilmanGlatzmaier81} and all subsequent studies.
In what follows we will only consider the linearised anelastic equations, so this term is absent.

As described in Appendix~\ref{sec: model-derivation},
in the LBR formulation
all thermodynamic quantities are expressed in terms of $p_1$ and $s_1$,
and the momentum is written in the form
\begin{align}
    \frac{\DD \mathbf{u}}{\DD t} + 2 \mathbf{\Omega} \times \mathbf{u} = \frac{s_1}{c_p} g \uvect{z} - \nablab \left(\frac{p_1}{p_0} \right) + \frac{1}{\mu_0 \rho_0} ( \nablab \times \mathbf{B} ) \times \mathbf{B}\,.
\end{align}
Moreover, the diabatic term, $Q$, is expressed solely in terms in terms of $s_1$, neglecting any contribution from $p_1$;
this is equivalent to adopting the approximate equation of state
\begin{equation}
  T_1 = \left( \frac{\partial T}{\partial s} \right)_p s_1
\end{equation}
in place of (\ref{eq:temp}).

How accurately the anelastic approximation captures magnetic buoyancy is unclear.
\citet{Fan01} verified that the anelastic model has the correct stability properties
in the asymptotic limit in which it is formally valid,
i.e., for $v \ll c$ and adiabatic stratification.
However, since \citeauthor{Fan01}'s analysis only considered the ideal equations
it does not differentiate between the GGG and LBR implementations.
\citet{Berkoff10} have compared these two formulations of the anelastic approximation, finding that while both produce consistent results for a background state of nearly uniform entropy,
they differ under more general conditions.

\subsection{Pseudo-incompressible}

The pseudo-incompressible model was originally derived by \citet{Durran89} as an improvement upon the anelastic model,
although it can also be viewed as a generalisation of earlier ``low Mach number'' models
to include stratification
\citep{RehmBaum78}.
Similar to the anelastic model, it is asymptotically valid in the limit where the background state is adiabatically stratified, but it retains terms that are formally negligible in this limit.
A detailed discussion on the asymptotic validity of the anelastic and pseudo-incompressible models can be found in \citet{Klein-etal10}.
A key feature of the pseudo-incompressible model is that the
velocity satisfies an inhomogeneous constraint of the form
\begin{equation}
  \nablab\cdot\mathbf{u} = \frac{g}{c^2}u_z
  - \left(\frac{\partial\rho}{\partial T}\right)_p \frac{Q}{\rho^2 c_p}
  \label{eq:PI-constraint}
\end{equation}
where we have used the same notation as in Section~\ref{sec: fully compressible}.
This means that the fluid expands in response to heating.
(In the anelastic model, by contrast, the expansion of fluid elements is dictated by the density of their surroundings, $\rho_0$.)

A generalisation of the pseudo-incompressible model that exactly obeys the laws of thermodynamics was presented by \citet{KleinPauluis12}.
Subsequently,
\citet{Vasil13} showed that the pseudo-incompressible approximation can be derived very efficiently using Lagrangian dynamics,
either by imposing a constraint on the fluid pressure
or, equivalently, by linearising the fluid action in the pressure variable.
This derivation can be generalised to include non-ideal physics \citep{Gay-Balmaz19}
and a time-dependent background state
\citep{SnodinWood22}.
\citeauthor{Vasil13} also obtained an MHD extension of the pseudo-incompressible model, by imposing a constraint on the total (i.e., gas plus magnetic) pressure.
The resulting momentum equation has the form
\begin{align}
    \rho \frac{\DD \mathbf{u}}{\DD t}
    + 2 \rho \, \mathbf{\Omega} \times \mathbf{u}
    &= - \nablab (\pi_0 + \pi_1) - g\rho\uvect{z} + \mathbf{B} \cdot \nablab \mathbf{H} + \frac{\pi_1\nablab\pi_0}{\rho(c^2+v^2)}\,,
    \label{eq:Vasil_mom} 
\end{align}
where $\pi_0(\mathbf{x})$ is
the total background pressure and where
\begin{equation}
  \mathbf{H} \equiv \left(1 + \frac{\pi_1}{\rho(c^2+v^2)} \right)\frac{\mathbf{B}}{\mu_0}\,.
\end{equation}
This must be solved
together with a complicated velocity constraint;
in the absence of diabatic terms (i.e., for $Q = 0$),
this constraint is
\begin{align}
    \rho(c^2+v^2) \nablab \cdot \mathbf{u} + \mathbf{u} \cdot \nablab \pi_0 &= \frac{1}{\mu_0} (\mathbf{B} \cdot \nablab \mathbf{u})\cdot \mathbf{B} \,.
\end{align}
In the absence of magnetic field (i.e., with $\mathbf{B} = \mathbf{0}$, $v=0$ and $\pi = p$)
these equations reduce to the pseudo-incompressible model of \citet{KleinPauluis12}.

The main advantage of this model is that, owing to its variational derivation,
it is guaranteed to conserve energy.
However, the complexity of the equations makes solving them prohibitive, and the asymmetric form of the Lorentz force in equation~(\ref{eq:Vasil_mom}) has no simple physical explanation.

The derivation performed by \citeauthor{Vasil13} does not make any assumption about the relative magnitudes of the sound speed, $c$, and the Alfv\'en speed, $v$,
but in the solar interior we expect that $v \ll c$.
In Appendix~\ref{sec: appendix-PI},
we show that this extra condition can be explicitly incorporated into the variational derivation,
ultimately leading to a pseudo-incompressible MHD model in which the Lorentz force retains its usual form, i.e.~we have the momentum equation:
\begin{align}
    \rho \frac{\DD\mathbf{u}}{\DD t} + 2 \rho \, \mathbf{\Omega} \times \mathbf{u} &= - \nablab (p_0 + p_1) - g \rho \uvect{z} + \frac{1}{\mu_0}(\nablab\times\mathbf{B})\times\mathbf{B} + \frac{p_1}{\rho c^2}\nablab p_0\,,
    \label{eq:momentum_PI} 
\end{align}
and the velocity constraint takes the same form as equation~(\ref{eq:PI-constraint}).
It is this version of the pseudo-incompressible equations that we will consider in our study.

\section{General Sound-proof Model \label{sec: sound-proof} }
To ascertain how well each of the sound-proof models discussed earlier describes magnetic buoyancy,
we consider a general linearised sound-proof model that includes each of the aforementioned sound-proof models as special cases.
This is a more efficient approach than simply analysing each individual model in turn.

Our general sound-proof model is given by the following set of linearised equations:
\begin{align}
    \rho_0 \frac{\partial}{\partial t}\mathbf{u} + 2 \rho_0 \mathbf{\Omega} \times \mathbf{u} &= - \left[\frac{D}{c^2}  p_1 + \left(\frac{\partial \rho}{\partial s} \right)_p s_1 \right]g \mathbf{\hat{z}} - \nablab \pi_1 + \frac{1}{\mu_0}\mathbf{B}_0\cdot\nablab\mathbf{B}_1 + \frac{1}{\mu_0} \mathbf{B}_1\cdot\nablab\mathbf{B}_0\,,
    \label{eq:sound-proof-momentum} \\
    \frac{\partial}{\partial t}\mathbf{B}_1 &= \nablab \times (\mathbf{u} \times \mathbf{B}_0)\,, \\
    \left( \frac{\partial \rho}{\partial s} \right)_p \frac{\partial}{\partial t}s_1 - \rho_0 H_s^{-1} u_z
    &= \left(\frac{\partial\rho}{\partial T}\right)_p \frac{Q_1}{\rho_0 c_p}\,,
    \label{eq:sound-proof-heat} \\
    \nablab \cdot \mathbf{u} &= C \frac{g}{c^2} u_z - F \left(\frac{\partial\rho}{\partial T}\right)_p \frac{Q_1}{\rho_0^2 c_p}\,,
    \label{eq:sound-proof-constraint} \\
    J\pi_1 &= p_1 + \frac{B_0}{\mu_0} B_{1x}\,, \\
    \left( \frac{\partial \rho}{\partial T} \right)_p T_1 &= \left( \frac{\partial \rho}{\partial s} \right)_p s_1 - G \frac{\gamma - 1}{c^2} p_1\,,
    \label{eq:sound-proof-temperature}
\end{align}
where we have used the same notation as for the fully compressible equations (\ref{eq:momentum})--(\ref{eq:EoS}). 
The coefficients $C$, $D$, $F$, $G$, $J$ are assumed to be known functions of altitude, $z$;
different choices for these coefficients correspond to different sound-proof approximations. 
For example, in the pseudo-incompressible approximation all of these coefficients are equal to $1$.
We will therefore assume that all the coefficients are of order unity.
Table \ref{table: coeff values} shows the values of these coefficients for each of the models discussed in the previous section.
Justification for the values in Table \ref{table: coeff values}, as well as more detailed derivations of each of the models, is provided in Appendix \ref{sec: model-derivation}.

Note that in our sound-proof model we have deliberately not included a density perturbation, $\rho_1$.
This is because there is often ambiguity in how density should be defined in a sound-proof system.
For example, \citet{Durran89} defined two quantities called $\rho$ and $\rho^\ast$, one that satisfies the equation of state and one that satisfies the continuity equation.
Note that in many implementations of the anelastic approximation \citep[e.g.,][]{OguraPhillips62,LippsHemler82,Lantz92,BraginskyRoberts95}
an expression for density is not explicitly needed because the equations are written in terms of pressure and entropy.
Combining equations~(\ref{eq:sound-proof-heat}) and (\ref{eq:sound-proof-constraint}),
we find that
\begin{align}
  \frac{\partial}{\partial t}F \left( \frac{\partial \rho}{\partial s} \right)_p s_1 =
  \left[F H_s^{-1} + C\frac{g}{c^2} - H_\rho^{-1}\right] \rho_0 u_z
  - \nablab\cdot(\rho_0\mathbf{u})\,,
\end{align}
which has the same form as the continuity equation~(\ref{eq:continuity})
if we define $\rho_1 \equiv F \left( \dfrac{\partial \rho}{\partial s} \right)_p s_1$
and if the coefficients $C$ and $F$ are chosen such that
\begin{equation}
  F H_s^{-1} + C\frac{g}{c^2} = H_\rho^{-1}\,.
  \label{eq:continuity_constraint}
\end{equation}
However, this definition of $\rho_1$ does not satisfy the equation of state~(\ref{eq:EoS}),
and nor, in general, is it the same quantity that appears in the buoyancy term of equation~(\ref{eq:sound-proof-momentum}).
For this reason, we will later refer to the quantity $\rho_\text{k} \equiv F \left( \dfrac{\partial \rho}{\partial s} \right)_p s_1$ as the ``kinematic density''.

\begin{table}[ht]
    \begin{tabular}{@{}cccccc@{}}
        \toprule 
        \toprule
        \diagbox{Coefficient}{Model} &
        \hspace{0.1cm} \thead{LBR \\ anelastic} \hspace{0.1cm} &
        \hspace{0.1cm} \thead{GGG \\ anelastic} \hspace{0.1cm} &
        \hspace{0.1cm} \thead{Pseudo-\\ incompressible} \hspace{0.1cm} &
        \hspace{0.1cm} \thead{Magneto-\\Boussinesq} \hspace{0.1cm} \\ \midrule
        $C$ & $\frac{c^2}{g} H_\rho^{-1}$ & $\frac{c^2}{g} H_\rho^{-1}$ & 1 & $\frac{c^2}{g} H_\rho^{-1}$ \\
        $D$ & $\frac{c^2}{g} H_\rho^{-1}$ & 1 & 1 & 1 \\
        $F$ & 0 & 0 & 1 & 0 \\
        $G$ & 0 & 1 & 1 & 1 \\
        $J$ & 1 & 1 & 1 & 0 \\ 
        \bottomrule
        \bottomrule
    \end{tabular}
    \caption{Values for each of the coefficients in our sound-proof model needed to reproduce the models summarised in Section \ref{sec: existing sound-proof}. 
    } 
    \label{table: coeff values}
\end{table}

\section{Non-rotating case \label{sec: non-rotating} }

Our goal now is to solve the linearised equations arising from the fully compressible and sound-proof models.
In both cases,
since the background state depends only on altitude, $z$, we can
seek solutions in the form $f_1 = \tilde{f}_1(z) \exp(\sigma t + \ii k_x x + \ii k_y y)$,
where $f$ represents any of the perturbed variables.
The linearised equations then reduce to a second-order system of linear ordinary differential equations (ODEs) in $z$.
In general this system cannot be solved analytically,
so to make progress we will consider several asymptotic limits of relevance to the interior of the Sun and other stars.

\subsection{Fast thermal relaxation \label{sub: infinite}}

We first consider the case with no rotation and $\alpha \to \infty$.
In this limit, the temperature perturbation vanishes \citep{Gilman70}.
The specific asymptotic regime we consider is given by
\begin{align*}
    v &\ll c\,, \\
    H_B &\sim H_\rho \sim H_s\,, \\
    k_y, \tfrac{\dd}{\dd z} &\gtrsim k_x, H_\rho^{-1}\,, \\
    \alpha &\to \infty\,, \\
    \sigma &\sim v/H_\rho\,.
\end{align*}
By adopting the scalings $k_y, \frac{\dd}{\dd z} \gtrsim k_x, H_\rho^{-1}$
we can initially assume that all length-scales are equal,
before subsequently assuming smaller length-scales in the directions perpendicular to the magnetic field, if desired.
The scaling for the growth rate, $\sigma$, is justified by the results, and is the expected timescale for magnetic buoyancy instability based on previous studies.
We apply this scaling regime to our fully compressible equations (\ref{eq:momentum})--(\ref{eq:EoS}) and retain only the leading-order terms.
The equations can then be reduced to a pair of coupled ODEs:
\begin{align}
    \left(\frac{\dd}{\dd z} + \frac{\gamma g}{c^2}\frac{\sigma^2}{q}\right)\pi_1 &= - \frac{1}{\sigma} \left(q + \gamma g \frac{v^2}{c^2} \left(\frac{\sigma^2}{q}H_\rho^{-1} - H_B^{-1}\right)\right) \rho_0 u_z\,, \label{eqn: fc nr if 1} \\
    \left(\frac{\dd}{\dd z} - H_\rho^{-1}\frac{\sigma^2}{q}\right)u_z &= - \frac{\sigma}{q} (k_x^2 + k_y^2) \frac{\pi_1}{\rho_0}\,, \label{eqn: fc nr if 2}
\end{align}
where we have defined
\begin{equation}
  q \equiv \sigma^2 + v^2k_x^2\,.
  \label{eq:q}
\end{equation}
Since we have neglected viscosity and magnetic diffusion,
the most unstable modes are found in the limit $k_y \to \infty$.
In this limit the total pressure perturbation, $\pi_1$,
and hence the left-hand side of equation~(\ref{eqn: fc nr if 1}),
becomes negligible,
and we are left with a local dispersion relation:
\begin{equation}
  (\sigma^2 + v^2k_x^2)^2 + \gamma g \frac{v^2}{c^2}H_B^{-1} \left(\sigma^2 \left(\frac{H_B}{H_\rho} - 1\right) - v^2k_x^2\right) = 0\,,
  \label{eq:diffusion_dispersion}
\end{equation}
which exactly matches the result of \citet{Gilman70}.
In the case of interchange motions (i.e.,~for $k_x=0$)
we have instability if and only if $H_B^{-1} > H_\rho^{-1}$.
But if $0 < H_B^{-1} < 2H_\rho^{-1}$ then the fastest-growing mode is undular, with $k_x^2 = \dfrac{\gamma g}{2c^2H_B}(1 - H_\rho/2H_B)$ and $\sigma^2 = \tfrac{1}{4}\gamma g(v^2/c^2)H_\rho/H_B^2\,$.

If we apply the same analysis to the general sound-proof model (\ref{eq:sound-proof-momentum})--(\ref{eq:sound-proof-temperature}),
we again eventually obtain a pair of coupled ODEs:
\begin{align}
    &\left( \frac{\dd}{\dd z} + (D + G (\gamma - 1)) \frac{g}{c^2} \left(J - \frac{v^2k_x^2}{q}\right) \right) \pi_1 = \nonumber \\
    &- \frac{1}{\sigma} \left(q + (D + G (\gamma - 1)) \frac{gv^2\sigma^2}{qc^2}\left(C\frac{g}{c^2} + F H_s^{-1} - \frac{q}{\sigma^2}H_B^{-1}\right)\right) \rho_0 u_z\, ,  \label{eqn: sp nr if 1}\\
    &\left(\frac{\dd}{\dd z} - \left(C\frac{g}{c^2} + F H_s^{-1}\right)\frac{\sigma^2}{q}\right)u_z = - \frac{\sigma}{q} (k_x^2 + k_y^2) \frac{\pi_1}{\rho_0}\, . \label{eqn: sp nr if 2}
\end{align}
Hence we can reproduce the fully compressible result
(\ref{eqn: fc nr if 1})--(\ref{eqn: fc nr if 2}) provided that
\begin{align}
    J &\simeq 1\,, \label{eq:infinite-1} \\
    D + G (\gamma - 1) &\simeq \gamma\,, \\
    C\frac{g}{c^2} + F H_s^{-1} &\simeq H_\rho^{-1}\,. \label{eq:infinite-3}
\end{align}
Note that we use ``$\simeq$'' here because we require these results to hold only at leading order,
for the particular asymptotic regime we have considered.

Referring to the particular sound-proof models listed in Table~\ref{table: coeff values},
we see immediately that the GGG anelastic model satisfies all of these constraints.
The same is true of the pseudo-incompressible model, when we recall that the final term in equation~(\ref{eq:g}) is negligible under our scaling assumptions.
However, the LBR anelastic model does not satisfy these constraints,
and therefore does not correctly describe the wavelength or growth rate of the magnetic buoyancy instability in this regime.
The magneto-Boussinesq approximation has $J=0$, and therefore does not satisfy the constraint (\ref{eq:infinite-1}). However, on
smaller scales in $y$ and/or $z$ (which was the regime considered by
\citet{SpiegelWeiss82})
the term involving $J$ on the left-hand side of equation~(\ref{eqn: sp nr if 1}) is negligible.  This is because, for small scales in $z$, the $z$-derivative term dominates the left-hand side of equation~(\ref{eqn: sp nr if 1}).
For small scales in $y$, the perturbation $\pi_1$ becomes vanishingly small, and so the entire left-hand side of equation~(\ref{eqn: sp nr if 1}) is negligible. 
In particular, the condition $J \simeq 1$ is not required in order to correctly reproduce the fastest growing mode (which, in the absence of viscosity and resistivity, has $k_y \to \infty$).

Interestingly, equation~(\ref{eq:infinite-3}) is exactly the result (\ref{eq:continuity_constraint})
required to achieve a form of mass conservation in the sound-proof model.
Here we have arrived at the same result on purely dynamical grounds,
without any explicit reference to mass conservation.

\subsection{No thermal relaxation \label{sub: zero}}

We now consider the ideal limit, with $\alpha = 0$.
We also reduce the strength of the (stabilising) thermal stratification,
which otherwise overwhelms the destabilising effect of magnetic buoyancy
in the absence of thermal relaxation.
Specifically, we now assume that $H_s^{-1} \sim (v^2/c^2) H_\rho^{-1} \ll H_\rho^{-1}$.
Note that this means the background state now has $g \simeq c^2/H_\rho$\,,
since both the $H_s$ and $H_B$ terms are negligible in the magneto-hydrostatic equation~(\ref{eq:g}).
The complete regime is given by
\begin{align*}
    v &\ll c\,, \\
    H_B &\sim H_\rho \sim (v^2/c^2)H_s\,, \\
    k_y, \tfrac{\dd}{\dd z} &\gtrsim k_x\,, H_\rho^{-1}\,, \\
    \sigma &\sim v/H_\rho\,, \\
    \alpha &= 0\,. 
\end{align*}
Again, we apply this scaling regime to the fully compressible equations (\ref{eq:momentum})--(\ref{eq:EoS}) and retain only the leading-order terms. This leads to
\begin{align}
    \left(\frac{\dd}{\dd z} + H_\rho^{-1}\frac{\sigma^2}{q}\right)\pi_1 &= - \frac{1}{\sigma} \left(q + c^2 H_\rho^{-1} H_s^{-1} + v^2 H_\rho^{-1} \left(\frac{\sigma^2}{q} H_\rho^{-1} - H_B^{-1} \right)\right) \rho_0 u_z\,, \label{eqn: fc nr nd 1} \\
    \left(\frac{\dd}{\dd z} - H_\rho^{-1}\frac{\sigma^2}{q}\right)u_z &= - \frac{\sigma}{q} \left( k_x^2 + k_y^2 \right) \frac{\pi_1}{\rho_0}\,. \label{eqn: fc nr nd 2}
\end{align}
The fastest growing mode is again found in the limit $k_y \to \infty$,
for which $\pi_1 \to 0$,
resulting in a local dispersion relation similar to (\ref{eq:diffusion_dispersion}):
\begin{equation}
(\sigma^2 + v^2 k_x^2)^2 + v^2 H_\rho^{-1} \left(\sigma^2 \left(H_\rho^{-1} - H_B^{-1} + \frac{c^2}{v^2} H_s^{-1} \right) + v^2 k_x^2 \left(\frac{c^2}{v^2} H_s^{-1} - H_B^{-1} \right) \right) = 0\,.
\label{eq:ideal_dispersion}
\end{equation}
We can directly translate the results from the previous section
by replacing $\frac{c^2}{\gamma} \to c^2 = gH_\rho$
and then replacing $H_B^{-1} \to H_B^{-1} - \frac{c^2}{v^2} H_s^{-1}$.
This shows that, in the absence of fast thermal relaxation,
the isothermal sound speed $c/\sqrt{\gamma}$ is replaced by the adiabatic sound speed $c$,
and the entropy gradient opposes the destabilising field gradient.

When we apply the same analysis to our sound-proof model (\ref{eq:sound-proof-momentum})--(\ref{eq:sound-proof-temperature}),
we obtain
\begin{align}
    \left( \frac{\dd}{\dd z} + D H_\rho^{-1} \left(J - \frac{v^2k_x^2}{q}\right) \right) \pi_1 &= - \frac{1}{\sigma} \left(q + c^2 H_\rho^{-1} H_s^{-1} + D v^2 H_\rho^{-1}\left( C \frac{\sigma^2}{q} H_\rho^{-1} - H_B^{-1}\right)\right) \rho_0 u_z\,,  \\
    \left(\frac{\dd}{\dd z} - CH_\rho^{-1}\frac{\sigma^2}{q}\right)u_z &= - \frac{\sigma}{q} (k_x^2 + k_y^2) \frac{\pi_1}{\rho_0}\,.
\end{align}
Hence we can reproduce the fully compressible result (\ref{eqn: fc nr nd 1})--(\ref{eqn: fc nr nd 2}) provided that 
\begin{align}
    J &\simeq 1\,, \label{eq:zero-1} \\
    D &\simeq 1\,, \\
    C &\simeq 1\,. \label{eq:zero-3}
\end{align}
The last of these constraints seems at first to be incompatible with what we found previously,
i.e.,~equation~(\ref{eq:infinite-3}),
but in the current asymptotic regime we have $H_s \gg H_\rho$ and $g \simeq c^2/H_\rho$.
Therefore (\ref{eq:zero-3}) is actually just a weaker version of the earlier constraint~(\ref{eq:infinite-3}).

\subsection{Finite $\alpha$}

The asymptotic regimes considered in Sections~\ref{sub: infinite}
and \ref{sub: zero} are easily generalised
to include a finite Newtonian cooling rate;
in particular by choosing
$\alpha \sim \sigma\frac{c^2}{v^2}\frac{H_\rho}{H_s}$
we retain the maximum number of terms in the leading-order equations.
In this way, we can obtain stronger constraints on the coefficients in the sound-proof model.
We do not present these results here, however, because
(a)~there is no straightforward analogy between a finite value of $\alpha$ and a finite thermal diffusivity $\kappa$,
and (b)~equally strong constraints can be obtained simply by combining the
results of Sections~\ref{sub: infinite} and \ref{sub: zero}.
Specifically, we have the constraints
$D = G = J = 1$ and $C\frac{g}{c^2} + F H_s^{-1} = H_\rho^{-1}$.
Taken together, these constraints remove four of the five degrees of freedom present in our general sound-proof model.

\section{Rotating Case \label{sec: rotating} }

The growth rate of the magnetic buoyancy instability, for parameters typical of the solar interior,
is comparable to or slower than the Sun's rotation rate.
We therefore expect the instability to be significantly affected by this rotation.
The main interest in this instability is to explain the appearance of active regions in the Sun at low latitudes,
where the rotation axis is roughly perpendicular to the direction of gravity,
and so in the following we will take $\mathbf{\Omega} = \Omega\uvect{y}$.
The presence of rotation generally acts to reduce the growth rate of the instability, and to separate the roots of the dispersion relation into ``fast'' and ``slow'' modes.
In the absence of magnetic diffusion,
we expect
that the slow (i.e.,~magnetostrophic) branch is most easily destabilised,
so we will only consider that branch here.
As in the previous section, we focus on the limits of infinite and zero thermal relaxation.

\subsection{Fast thermal relaxation}

As with the non-rotating case, we first consider the limit $\alpha \to \infty$.
Our complete asymptotic regime is given by
\begin{align*}
   \Omega &\sim \left(\frac{c}{v}\right)^{1/2} v H_\rho^{-1}\,, \\
    v &\ll c\,, \\
    H_B &\sim H_\rho \sim H_s\,, \\
    k_y\,, \tfrac{\dd}{\dd z} &\gtrsim k_x\,, H_\rho^{-1}\,, \\
    \sigma &\sim \left( \frac{v}{c} \right)^{1/2} v H_\rho^{-1}\,, \\
    \alpha &\to \infty\,.
\end{align*}
Note that the growth rate is slower than in the nonrotating case,
and is of order $\sigma \sim v^2k_x^2/\Omega$,
characteristic of magnetostrophic dynamics.

We apply these scalings to the fully compressible equations (\ref{eq:momentum})--(\ref{eq:EoS}) and keep only the leading-order terms,
finding that
\begin{align}
    \left( \frac{\dd}{\dd z} - \frac{2 \Omega \sigma}{\ii v^2 k_x} \right) \pi_1 &= - \frac{1}{\sigma} \Bigg(v^2 k_x^2 - g \gamma \frac{v^2}{c^2} \left( H_B^{-1} + \frac{2 \Omega \sigma}{\ii v^2 k_x} \right) - \frac{2 \Omega \sigma}{\ii k_x} \left(H_\rho^{-1} + \frac{2 \Omega \sigma}{\ii v^2 k_x} \right) \Bigg) \rho_0 u_z\,, \\
    \bigg( \frac{\dd}{\dd z} + \frac{2 \Omega \sigma}{\ii v^2 k_x} \bigg) u_z &= - \frac{\sigma}{v^2} \left(1 + \frac{k_y^2}{k_x^2} \right) \frac{\pi_1}{\rho_0}\,.
\end{align}
The most unstable modes occur in the limit 
$k_y \to \infty$, for which we have the local dispersion relation
\begin{align}
     \frac{4 \Omega^2 \sigma^2}{v^2 k_x^2} - \frac{2 \Omega \sigma}{\ii k_x} \left( H_\rho^{-1} + \gamma \frac{g}{c^2} \right) - g \gamma \frac{v^2}{c^2}  H_B^{-1} + v^2 k_x^2 = 0\,.
     \label{eqn: local dispersion fast diffusion}
\end{align}
We have an (oscillatory) instability if
$H_B^{-1} > \frac{c^2}{4 \gamma g} \left( \gamma \frac{g}{c^2} + H_\rho^{-1} \right)^2$,
and the fastest growing mode then has
$k_x^2 = \frac{1}{2} \gamma \frac{g}{c^2} H_B^{-1} - \frac{1}{8} \left( \gamma \frac{g}{c^2} + H_\rho^{-1} \right)^2$.

Applying the same scaling regime to our sound-proof model (\ref{eq:sound-proof-momentum})--(\ref{eq:sound-proof-temperature}),
we obtain
\begin{align}
    &\left(\frac{\dd}{\dd z} + \bigl(D + G(\gamma - 1)\bigr) \frac{g}{c^2} (J - 1) - \frac{2 \Omega \sigma}{\ii v^2 k_x} \right) \pi_1 + \frac{1}{\sigma} \bigg( - \bigl(D + G(\gamma - 1)\bigr) \frac{g}{c^2} v^2 \left(H_B^{-1} + \frac{2 \Omega \sigma}{\ii v^2 k_x} \right) \nonumber \\
    &+ v^2 k_x^2 - \frac{2 \Omega \sigma}{\ii k_x} \left(\frac{2 \Omega \sigma}{\ii v^2 k_x} + C \frac{g}{c^2} + F H_s^{-1} \right) \bigg) \rho_0 u_z = 0\,,  \\
    &\left( \frac{\dd}{\dd z} + \frac{2 \Omega \sigma}{\ii v^2 k_x} \right) u_z = - \frac{\sigma}{v^2} \left(1 + \frac{k_y^2}{k_x^2} \right) \frac{\pi_1}{\rho_0}\,.
\end{align}
Comparing this with the fully compressible result, 
we recover the same constraints as in the non-rotating case,
i.e.,~equations~(\ref{eq:infinite-1})--(\ref{eq:infinite-3}), with the same conclusions regarding the applicability of the various sound-proof models in this regime.

\subsection{No thermal relaxation}

As for the non-rotating case, we now consider the ideal limit $\alpha = 0$ and we simultaneously change the scaling of $H_s^{-1}$ accordingly. 
The regime considered here is given by
\begin{align*}
   \Omega &\sim \left(\frac{c}{v}\right)^{1/2} v H_\rho^{-1}\,, \\
    v &\ll c\,, \\
    H_B &\sim H_\rho \sim (v^2 / c^2) H_s\,, \\
    k_y\,, \tfrac{\dd}{\dd z} &\gtrsim k_x\,, H_\rho^{-1}\,, \\
    \sigma &\sim \left( \frac{v}{c} \right)^{1/2} v H_\rho^{-1}\,, \\
    \alpha &= 0\,.
\end{align*}
In this regime the fully compressible equations (\ref{eq:momentum})--(\ref{eq:EoS}) reduce to the following pair of ODEs, where we have only kept the leading-order terms:
\begin{align}
    \left( \frac{\dd}{\dd z} - \frac{2 \Omega \sigma}{\ii v^2 k_x} \right) \pi_1 &= - \frac{1}{\sigma} \Bigg(v^2 k_x^2 + c^2 H_\rho^{-1} H_s^{-1} - \frac{2 \Omega \sigma}{\ii k_x} \left(H_\rho^{-1} + \frac{2 \Omega \sigma}{\ii v^2 k_x} \right) \nonumber \\
    &- v^2 H_\rho^{-1} \left( H_B^{-1} + \frac{2 \Omega \sigma}{\ii v^2 k_x} \right)\Bigg) \rho_0 u_z\,,\\
    \left( \frac{\dd}{\dd z} + \frac{2 \Omega \sigma}{\ii v^2 k_x} \right) u_z &= - \frac{\sigma}{v^2} \left(1 + \frac{k_y^2}{k_x^2} \right) \frac{\pi_1}{\rho_0}\,.
\end{align}
Again, the most unstable mode arises
in the limit $k_y \to \infty$.
In this limit $\pi_1 \to 0$, and we obtain the local dispersion relation
\begin{align}
    \frac{4 \Omega^2 \sigma^2}{v^2 k_x^2} - \frac{4 \Omega \sigma}{\ii k_x} H_\rho^{-1} - v^2 H_\rho^{-1} H_B^{-1} + c^2 H_\rho^{-1} H_s^{-1} + v^2 k_x^2 = 0\,.
\end{align}
As in the non-rotating case,
we can directly translate the results from the previous section, equation (\ref{eqn: local dispersion fast diffusion}),
by replacing
$\frac{c^2}{\gamma} \to c^2 = gH_\rho$,
and then replacing
$H_B^{-1} \to H_B^{-1} - \frac{c^2}{v^2} H_s^{-1}$.

Applying the same analysis to our sound-proof model (\ref{eq:sound-proof-momentum})--(\ref{eq:sound-proof-temperature}), we obtain
\begin{align}
    \left(\frac{\dd}{\dd z} + D H_\rho^{-1} (J - 1) - \frac{2 \Omega \sigma}{\ii v^2 k_x} \right) \pi_1 &= - \frac{1}{\sigma} \Bigg( c^2 H_\rho^{-1} H_s^{-1} - \frac{2 \Omega \sigma}{\ii k_x} \left(\frac{2 \Omega \sigma}{\ii v^2 k_x} + C H_\rho^{-1} \right) \nonumber \\
    &- D H_\rho^{-1} v^2 \Bigg( H_B^{-1} + \frac{2 \Omega \sigma}{\ii v^2 k_x} \Bigg) + v^2 k_x^2 \Bigg) \rho_0 u_z\,, \\
    \left( \frac{\dd}{\dd z} + \frac{2 \Omega \sigma}{\ii v^2 k_x} \right) u_z &= - \frac{\sigma}{v^2} \left(1 + \frac{k_y^2}{k_x^2} \right) \frac{\pi_1}{\rho_0}\,. 
\end{align}
Comparing this with the fully compressible result
we recover the same constraints as in the non-rotating case,
i.e.,~equations~(\ref{eq:zero-1})--(\ref{eq:zero-3}).

\section{Conclusions \label{sec: conclusion} }

We have introduced a very general sound-proof MHD model,
and constrained the coefficients in the model by considering the linear onset of magnetic buoyancy instability under a range of physical conditions.
The most general model that satisfies all of these constraints has the form
\begin{align}
    \rho_0 \frac{\partial}{\partial t}\mathbf{u} + 2 \rho_0 \mathbf{\Omega} \times \mathbf{u} &= \rho_\text{d}\mathbf{g} - \nablab \pi_1 + \frac{1}{\mu_0}\mathbf{B}_0\cdot\nablab\mathbf{B}_1 + \frac{1}{\mu_0} \mathbf{B}_1\cdot\nablab\mathbf{B}_0\,,
    \label{eq: momentum conclusion}
    \\
    \frac{\partial}{\partial t}\mathbf{B}_1 &= \nablab \times (\mathbf{u} \times \mathbf{B}_0)\,, \\
    \frac{\partial}{\partial t}\rho_\text{k} &= - \nablab\cdot(\rho_0\mathbf{u})\,, \label{eq: continuity conclusion}
    \\
    \frac{\partial}{\partial t}s_1 + \mathbf{u}\cdot\nablab s_0 &= \frac{Q_1}{\rho_0 T_0}\,, \label{eq: energy conclusion}
    \\
\pi_1 &= p_1 + \frac{1}{\mu_0}\mathbf{B}_0\cdot\mathbf{B}_1\,,
    \label{eq: pressure conclusion}
\end{align}
where $\rho_\text{d} \equiv \frac{1}{c^2}p_1 + \left( \frac{\partial \rho}{\partial s}\right)_p s_1$
and $\rho_\text{k} \propto s_1$.
We call $\rho_\text{d}$ the ``dynamic density'' because it determines the buoyancy force in equation~(\ref{eq: momentum conclusion}); it is related to $p_1$ and $s_1$ by the usual equation of state.
We call $\rho_\text{k}$ the ``kinematic density''
because it satisfies the continuity equation~(\ref{eq: continuity conclusion});
it is simply proportional to $s_1$, with a coefficient that is essentially arbitrary --- using the notation introduced in Section~\ref{sec: sound-proof} we have $\rho_\text{k} = F \left( \frac{\partial \rho}{\partial s} \right)_p s_1$ where $F$ remains unconstrained.
Since $\rho_k$ is proportional to $s_1$, equations~(\ref{eq: continuity conclusion}) and (\ref{eq: energy conclusion}) are not independent,
and in order for them to both be satisfied we must have 
\[
  \nablab \cdot \mathbf{u} = [H_\rho^{-1} - F H_s^{-1}] u_z
   - F \left(\frac{\partial\rho}{\partial T}\right)_p \frac{Q_1}{\rho_0^2 c_p} \,,
\]
which precisely corresponds to our model's velocity constraint (\ref{eq:sound-proof-constraint})
under the condition that $C\frac{g}{c^2} + F H_s^{-1} = H_\rho^{-1}$.

Beyond the specific parameter regimes presented here,
we have also considered a finite cooling rate $\alpha$, a tilted rotation axis,
and the fast branch of the rotating dispersion relation.
In all of these cases,
we find that the model given by (\ref{eq: momentum conclusion})--(\ref{eq: pressure conclusion})
captures the correct leading-order behaviour.

We emphasize that some of the parameter regimes we have considered lie outside the asymptotic regimes in which sound-proof models are usually derived.
In particular, the magneto-Boussinesq, anelastic and pseudo-incompressible models are usually derived under the assumption that the fluid has nearly uniform entropy (in the sense that $H_s \gg H_\rho$).
As expected,
the anelastic and pseudo-incompressible models correctly describe the linear onset of the instability
when $H_s \gg H_\rho$,
and
the magneto-Boussinesq approximation is applicable on small scales (in the directions perpendicular to the magnetic field).
These results are consistent with those of \citet{Wilczynski22},
who performed a similar analysis of the anelastic model,
but restricted attention to regimes in which it is asymptotically valid.
Only the GGG~anelastic and pseudo-incompressible approximations predict the instability onset correctly in all of the regimes that we have considered,
because both can be written in exactly the form of equations~(\ref{eq: momentum conclusion})--(\ref{eq: pressure conclusion}),
with $F=0$ and $F=1$, respectively.

In this study we have assessed different linearised sound-proof models entirely on the basis of whether they accurately describe the magnetic buoyancy instability.
However, there are other important considerations when choosing between the various models.
For example, it is known that some (non-magnetic) sound-proof models have a Hamiltonian structure,
implying that both the nonlinear and linearized equations conserve a form of energy
\citep{Bernardet95,Bannon96,Brown12,Vasil13}.
We show in Appendix~\ref{sec: appendix-PI} that the MHD pseudo-incompressible model
has a Hamiltonian structure,
and therefore also conserves energy.
This is a beneficial property for any model, because it can be used to establish stability criteria \citep{Bernstein58} and to rule out certain unphysical behaviours \citep[e.g.][]{Jones09}.
A thorough examination of energy conservation in MHD sound-proof models is beyond the scope of this paper, but will be presented in a later publication.

Another important consideration when comparing numerical models is their computational complexity.
The main motivation for using a sound-proof model is the reduced computational burden in comparison with a fully compressible mode.
However, within the set of sound-proof models,
some are more computationally expensive than others.
In particular, as described in Section \ref{sec: existing sound-proof}, a significant advantage of the LBR anelastic model is that the pressure and temperature perturbations do not need to be calculated explicitly. 
In the GGG and pseudo-incompressible models,
by contrast, the pressure perturbation
needs to be calculated explicitly by solving an elliptic equation
\citep{Bernardet95,Bannon06}.
Moreover, in the pseudo-incompressible model the velocity field satisfies an inhomogeneous constraint that must be solved in tandem with the pressure perturbation \citep{SnodinWood22}.
Interestingly, our results show that the models that best reproduce the linear behaviour of the fully compressible system are also the models that are more 
computationally expensive 
to solve.

Although we have here only considered the linear instability problem,
magnetic buoyancy in the Sun is certainly a nonlinear process,
and so full understanding of the formation of active regions can only come from a nonlinear model.
However, it seems likely that a necessary condition for accurately describing the full, nonlinear problem would be to accurately describe the linear regime. Hence, the work presented here is a necessary first step towards a nonlinear sound-proof model of magnetic buoyancy.
In nonlinear modelling energy budgets are of central importance,
and will be considered in subsequent work.

\begin{acknowledgments}
This work was supported by EPSRC grants EP/R024952/1 and EP/R51309X/1,
and by Leverhulme Trust grant RPG-2020-109.
\end{acknowledgments}

\appendix

\section{Asymptotic Derivation of Sound-Proof Models} \label{sec: model-derivation}

Here we summarize the assumptions under which the anelastic and Boussinesq approximations are usually derived.
For simplicity, and to facilitate comparison with the content of the paper, we consider only the linearised equations, and we neglect rotation.
We begin with the linearised fully-compressible equations
\begin{align}
    \rho_0 \frac{\partial}{\partial t}\mathbf{u} &= - g \rho_1 \uvect{z} - \nablab \pi_1 + \frac{1}{\mu_0}\mathbf{B}_0\cdot\nablab\mathbf{B}_1 + \frac{1}{\mu_0} \mathbf{B}_1\cdot\nablab\mathbf{B}_0\,,
    \label{eq:momentum_appendix} \\
\frac{\partial}{\partial t}\rho_1 &= - \nablab \cdot (\rho_0 \mathbf{u})\,,
    \label{eq:continuity_appendix} \\
    \frac{\partial}{\partial t}\mathbf{B}_1 &= \nablab \times (\mathbf{u} \times \mathbf{B}_0)\,, \label{eq:induction_appendix} \\
        \left( \frac{\partial \rho}{\partial s} \right)_p \frac{\partial}{\partial t}s_1 - \rho_0 H_s^{-1} u_z
    &= \left(\frac{\partial\rho}{\partial T}\right)_p \frac{Q_1}{\rho_0 c_p}\,,
    \label{eq:energy_appendix} \\
    \pi_1 &= p_1 + \frac{1}{\mu_0} \mathbf{B}_0\cdot\mathbf{B}_1\,, \\
    \left( \frac{\partial \rho}{\partial T} \right)_p T_1 &= \left( \frac{\partial \rho}{\partial s} \right)_p s_1 - \frac{\gamma - 1}{c^2} p_1\,, \label{eq:temp_appendix} \\
    \rho_1 &= \left( \frac{\partial \rho}{\partial s} \right)_p s_1 + \frac{1}{c^2} p_1\,,
    \label{eq:EoS_appendix}
\end{align}
where we recall that $g$ is related to the scale heights by magneto-hydrostatic balance:
\begin{equation}
  g = c^2 H_\rho^{-1} - c^2H_s^{-1} + v^2H_B^{-1}\,.
  \label{eq:gravity_appendix}
\end{equation}

\subsection{The GGG anelastic model} \label{sec: appendix-GGGanelastic}
In the GGG anelastic model,
all (non-magnetic) terms are retained apart from the left-hand side of the continuity equation~(\ref{eq:continuity_appendix}),
which therefore reduces to the ``anelastic equation''
\begin{align}
    \nablab \cdot (\rho_0 \mathbf{u}) = 0\,. \label{eq:anelastic_appendix}
\end{align}
This approximation can be justified by assuming that the perturbations have a length-scale that is comparable to the density scale height, $H_\rho$, and a timescale, $\tau$, that is much longer than the acoustic timescale, $H_\rho/c$.
Under these assumptions,
and assuming that the buoyancy force is of the same order as the fluid acceleration in equation~(\ref{eq:momentum_appendix}),
the left-hand side of equation~(\ref{eq:continuity_appendix})
is found to be of order $M^2$, where $M \equiv H_\rho/(c\tau) \ll 1$ is the Mach number.
However, in order for both terms on the left-hand side of equation~(\ref{eq:energy_appendix}) to be of the same order,
the timescale must be of order $\tau \sim \sqrt{H_s/g} \equiv 1/N$, where $N$ is the buoyancy frequency.
For consistency, we therefore require that
$H_s \sim H_\rho/M^2 \gg H_\rho$,
i.e.,~the background state must have nearly uniform entropy.
Moreover, in the anelastic approximation the background state is usually taken to be non-magnetic,
with the entire magnetic field regarded as a small perturbation.
This requires that the total magnetic pressure force is negligible in comparison with gravity,
i.e.,~that $g \gg v^2/H_B$.
Therefore, at leading order, equation~(\ref{eq:gravity_appendix})
becomes simply
\[
  g = c^2/H_\rho\,.
\]
Although the anelastic approximation is formally valid only when $H_s \gg H_\rho$,
it is quite often used in situations where this condition does not apply (which partly motivates the present study).

With the left-hand side of equation~(\ref{eq:continuity_appendix}) neglected,
equations~(\ref{eq:momentum_appendix})--(\ref{eq:EoS_appendix}) are equivalent to our general sound-proof model (\ref{eq:sound-proof-momentum})--(\ref{eq:sound-proof-temperature})
if we identify $C= \frac{c^2}{g} H_\rho^{-1}$, $F=0$, $D = G = J = 1$.
In particular, equation~(\ref{eq:anelastic_appendix})
can be written as
\[
  \nablab\cdot\mathbf{u} = u_z/H_\rho\,,
\]
which is equivalent to (\ref{eq:sound-proof-constraint}) when we set $C = \frac{c^2}{g} H_\rho^{-1}$ and $F=0$.

\subsection{The LBR anelastic model} \label{sec: appendix-LBRanelastic}
Without approximation, the pressure and gravity terms in the momentum equation~(\ref{eq:momentum_appendix}) can be re-written as 
\begin{align}
    - \rho_1 g \uvect{z} - \nablab p_1 = - \nablab \left( \frac{p_1}{\rho_0} \right) + \frac{1}{\rho_0} \left[ H_\rho^{-1} - \frac{g}{c^2} \right]p_1 \uvect{z} - \frac{1}{\rho_0} \left( \frac{\partial \rho}{\partial s} \right)_p s_1 g \uvect{z}\,. \label{eq: LBR appendix}
\end{align}
Under the conditions inherent in the anelastic approximation,
the term involving $\left[ H_\rho^{-1} - \frac{g}{c^2} \right]$
is negligible.
With this term neglected, the equations have a very similar mathematical form to the Boussinesq equations,
except that $\rho_0$ is not constant.
In particular,
in the absence of diabatic processes
(i.e.,~with $Q_1=0$)
the pressure perturbation $p_1$ appears only in the gradient term in the momentum equation, meaning that it does not need to be calculated explicitly.
However, this analogy with the Boussinesq approximation breaks down when thermal relaxation is included (i.e.,~with $Q_1\neq0$),
because the temperature perturbation, $T_1$, depends on $p_1$
via equation~(\ref{eq:temp_appendix}).
If the $p_1$ term in equation~(\ref{eq:temp_appendix}) is neglected,
the analogy with the Boussinesq equations is restored.
This point was independently discovered by 
\citet{Lantz92} and \citet{BraginskyRoberts95},
so we refer to the resulting equations as the LBR approximation.
Whereas \citeauthor{Lantz92} considered this merely as a mathematical convenience,
\citeauthor{BraginskyRoberts95} justified it by arguing that,
in a convective system, it is entropy gradients rather than temperature gradients that drive the flow of heat;
in this sense, omitting the $p_1$ term in equation~(\ref{eq:temp_appendix}),
and thus replacing temperature diffusion with entropy diffusion,
can be regarded as a mean-field prescription for the heat transport by small-scale convection.
More recently, \citet{Pauluis08} has shown that this approximation is also necessary for ``thermodynamic consistency'',
i.e.,~to make the anelastic equations consistent with the laws of thermodynamics.

In fact, a system of equations mathematically identical to the LBR model was earlier obtained by \citet{LippsHemler82}
under slightly different assumptions.
Specifically, they considered perturbations with a length-scale
$\ll H_\rho \sim H_s$,
but retained some next-to-leading-order terms in the resulting equations.
Under these conditions it can be shown that the $p_1$ term in equation~(\ref{eq:temp_appendix}) is negligible,
much as in the Boussinesq approximation.

If the anelastic approximation is used in circumstances where the background state is not close to adiabatic,
then the neglect of the $\left[ H_\rho^{-1} - \frac{g}{c^2} \right]$ term in equation (\ref{eq: LBR appendix}) makes a material difference to the results.
For this reason, the coefficient $D$ in Table~\ref{table: coeff values} takes different values for the GGG and LBR models.
Neglecting the $p_1$ term in equation (\ref{eq:temp_appendix}) is equivalent to setting $G=0$ in our general sound-proof model.

\subsection{The magneto-Boussinesq model} \label{subsection: magneto-Boussinesq}

The magneto-Boussinesq model was first discussed by \citet{SpiegelWeiss82},
and subsequently derived more rigorously by \citet{Corfield84}.
We follow \citeauthor{Corfield84}'s derivation by introducing two small parameters, $\epsilon_1$ and $\epsilon_2$,
and assuming the following hierarchy of length-scales:
\[
  \epsilon_1 \nablab_\perp \sim k_x \sim H_\rho^{-1} \sim H_B^{-1} \sim \frac{\epsilon_1}{\epsilon_2} H_s^{-1},
\]
where the subscript $\perp$ refers to the components perpendicular to the magnetic field $\mathbf{B}_0$.
We also assume that the
timescale is set by the buoyancy frequency, i.e.,
\[
  \frac{\partial}{\partial t} \sim N = \sqrt{g H_s^{-1}}\,,
\]
and that the
ratio of Alfv\'en speed to sound speed satisfies 
\[
  \frac{v^2}{c^2} \sim \frac{\epsilon_2}{\epsilon_1}\,.
\]
The magneto-Boussinesq model is obtained by adopting the ordering $\epsilon_2 \ll \epsilon_1 \ll 1$,
which implies that $v^2 \ll c^2$.

With these scalings in mind we can re-write the continuity equation~(\ref{eq:continuity_appendix})
and the induction equation~(\ref{eq:induction_appendix}) as
\begin{align}
    \underbrace{\nablab_\perp \cdot \mathbf{u}_\perp}_{O(1)}
    &=
    \underbrace{H_\rho^{-1} u_z - \ii k_x u_x}_{O(\epsilon_1)}
    + \underbrace{\frac{1}{\rho_0} \frac{\partial \rho_1}{\partial t}}_{O(\epsilon_2)}\,,
    \label{eq:magneto-mass} \\
    \underbrace{(\nablab_\perp \cdot \mathbf{u}_\perp) \mathbf{B}_0}_{O(1)}
    &=
    \underbrace{(\mathbf{B}_0 \cdot \nablab) \mathbf{u}
    + (H_B^{-1}u_z - \ii k_x u_x) \mathbf{B}_0
    - \frac{\partial\mathbf{B}_1}{\partial t}}_{O(\epsilon_1)}\,,
    \label{eq:magneto-induction}
\end{align}
where the relative magnitudes of $\mathbf{u}$, $\rho_1$ and $\mathbf{B}_1$ are inferred from the momentum equation~(\ref{eq:momentum_appendix}).
At leading order, these two equations contain the same information, and so it is necessary to consider the terms of order $\epsilon_1$.
This can be done by eliminating the $O(1)$ terms, resulting in
\begin{align*}
  \nablab_\perp \cdot \mathbf{u}_\perp &= 0, \\
  \frac{\partial\mathbf{B}_1}{\partial t} &=
  (\mathbf{B}_0 \cdot \nablab) \mathbf{u}
    + (H_B^{-1} - H_\rho^{-1}) u_z \mathbf{B}_0.
\end{align*}
Therefore the fields $\mathbf{u}_\perp$ and $\mathbf{B}_{1\perp}$ are solenoidal under this approximation,
but not the full fields $\mathbf{u}$ and $\mathbf{B}_1$.
This inherent anisotropy makes the magneto-Boussinesq approximation
unsuitable for many practical applications,
and so in our general sound-proof model we have retained all components of $\mathbf{u}$ and $\mathbf{B}_1$ in the velocity constraint and induction equation.
In order to still replicate the magneto-Boussinesq model as closely as possible, we retain the $O(1)$ and $O(\epsilon_1)$ terms in equations~(\ref{eq:magneto-mass}) and (\ref{eq:magneto-induction}),
which is equivalent to setting $C = \frac{c^2}{g}H_\rho^{-1}$ and $F = 0$.

In order for the pressure term $\nablab\pi$ in the momentum equation~(\ref{eq:momentum_appendix})
to be of the same order as the other terms,
the total pressure perturbation, $\pi_1$, must be smaller than
the perturbations of gas pressure, $p_1$,
and magnetic pressure, $\frac{1}{\mu_0}\mathbf{B}_0\cdot\mathbf{B}_1$,
by a factor of $\epsilon_1$.
In other words, these two contributions to the total pressure must cancel at leading order, with
\[
  p_1 + \frac{1}{\mu_0}\mathbf{B}_0\cdot\mathbf{B}_1 = 0.
\]
This means that $\pi_1$ simply becomes a Lagrange multiplier in the magneto-Boussinesq model,
and is mathematically independent of the other perturbed quantities.
In our general sound-proof model, this corresponds to setting $J=0$.

\section{Variational Derivation of MHD Pseudo-Incompressible Model}\label{sec: appendix-PI}

To our knowledge, there has not been a formal (i.e.,~rigorous asymptotic) derivation of the pseudo-incompressible model that includes magnetic fields.
An MHD version has been derived by \citet{Vasil13} using variational methods,
but this resulted in a non-standard form of the Lorentz force.
The derivation of \citet{Vasil13} assumed that perturbations to the total magnetic pressure remain small, but made no explicit assumption about the magnitude of the Alfv\'en speed, $v$, relative to the sound speed, $c$.
Here we will present a variational derivation of the MHD pseudo-incompressible model in which the smallness of $v/c$ is used to further simplify the action.
By making use of variational methods we necessarily neglect any diabatic processes, such as diffusion.
For simplicity we will also neglect rotation.

We begin with the action for a fully compressible MHD fluid:
\begin{equation}
  \mathcal{S} = \iint\mathcal{L}\,\dd\mathbf{x}^3\,\dd t\,,
\end{equation}
where the Lagrangian density, $\mathcal{L}$, can be written most compactly as
\begin{equation}
  \mathcal{L} = \tfrac{1}{2}\rho|\mathbf{u}|^2 - \rho\Phi - \rho U(\rho,s) - \frac{|\mathbf{B}|^2}{2\mu_0}\,.
\end{equation}
Here, $\Phi$ is the gravitational potential and $U(\rho,s)$ is the specific internal energy,
expressed in terms of its natural variables, $\rho$ and $s$.
The equations of motion can be derived by applying Hamilton's principle,
i.e.,~by considering the first variation of the action:
\begin{equation}
  \delta\mathcal{S} = \iint\left[(\tfrac{1}{2}|\mathbf{u}|^2 - \Phi - U - p/\rho)\delta\rho + \rho\mathbf{u}\cdot\delta\mathbf{u} - \rho T\delta s - \frac{1}{\mu_0}\mathbf{B}\cdot\delta\mathbf{B}\right]\,\dd^3\mathbf{x}\,\dd t
\end{equation}
where the pressure, $p$, and temperature, $T$, are defined via the usual equation of state:
\begin{equation}
  p \equiv \rho^2\frac{\partial U}{\partial\rho}
  \qquad \mbox{and} \qquad
  T \equiv \frac{\partial U}{\partial s}\,.
\end{equation}
The variations $\mathbf{u}$, $\rho$, $s$ and $\mathbf{B}$ can be expressed in terms of the Lagrangian displacement, $\boldsymbol{\xi}$, as \citep{Newcomb61} 
\begin{equation}
  \delta\rho = - \nablab\cdot(\rho\boldsymbol{\xi})\,, \quad
  \delta s = - \boldsymbol{\xi}\cdot\nablab s\,, \quad
  \delta\mathbf{u} = \frac{\partial\boldsymbol{\xi}}{\partial t} + \mathbf{u}\cdot\nablab\boldsymbol{\xi} - \boldsymbol{\xi}\cdot\nablab\mathbf{u}\,, \quad
  \delta\mathbf{B} = \nablab\times(\boldsymbol{\xi}\times\mathbf{B})\,.
  \label{eq:xi}
\end{equation}
By requiring that $\delta\mathcal{S}$ vanishes for all possible choices of $\boldsymbol{\xi}$, we thus eventually obtain the usual equation of motion for a compressible fluid, i.e., equation~(\ref{eq:full_momentum}), though without the Coriolis term.

\citet{Vasil13} showed that,
in the non-magnetic case,
the pseudo-incompressible model can be derived
by first making a Legendre transformation from internal energy, $U(\rho,s)$,
to enthalpy, $H(p,s) \equiv U + p/\rho$, while introducing the pressure, $p \equiv \rho^2\,\partial U/\partial\rho$,
as an additional independent variable,
and then linearising the action about a fixed reference pressure, $p_0(\mathbf{x})$.
To generalise this argument to a magnetised fluid, we must first recognise that it is perturbations to the \emph{total} pressure, $\pi$, that are expected to be small.
We therefore begin by introducing $\pi$ as an additional variable in the action,
by making an appropriate Legendre transformation.
Generalising the thermodynamic definition of $p$, we can write
\begin{equation}
  \pi \equiv \rho^2\frac{\partial \tilde{U}}{\partial\rho}\,,
\end{equation}
where $\tilde{U}(\rho,s,s_B)$ is the effective internal energy,
\begin{equation}
  \tilde{U}(\rho,s,s_B) = U(\rho,s) + \rho\,s_B\,,
\end{equation}
and $s_B \equiv \dfrac{|\mathbf{B}|^2}{2\mu_0\rho^2}$ is the ``magnetic entropy''.
The second derivative of $\tilde{U}$ with respect to $1/\rho$ is readily found to be $\rho^2(c^2+v^2)$, which is strictly positive,
and so $\tilde{U}$ is a convex function of $1/\rho$.
We can therefore define the effective enthalpy via the Legendre transformation
\begin{equation}
  \tilde{H}(\pi,s,s_B) \equiv \tilde{U}(\rho,s,s_B) + \pi/\rho\,,
\end{equation}
by analogy with the non-magnetic case.
We note that, in general, it is not possible to express the function $\tilde{H}(\pi,s,s_B)$ analytically, even when the equation of state $U(\rho,s)$ is known,
but nonetheless this function is well-defined.
With these definitions, we can now write the Lagrangian density as
\begin{equation}
  \mathcal{L} = \tfrac{1}{2}\rho|\mathbf{u}|^2 - \rho\Phi - \rho \tilde{H}(\pi,s,s_B) + \pi\,.
  \label{eq:action_exact}
\end{equation}
We emphasize that, up to this point, no approximation has been made;
if we apply Hamilton's principle to this action,
regarding $\pi$ as an independent variable,
then we eventually obtain the fully compressible equation of motion
as well as the equation of state
\begin{equation}
  \frac{1}{\rho} = \frac{\partial\tilde{H}}{\partial\pi}\,.
\end{equation}
We can now reproduce the pseudo-incompressible MHD model of \citet{Vasil13}
by writing the total pressure in equation~(\ref{eq:action_exact}) as $\pi = \pi_0(\mathbf{x}) + \pi_1$,
where $\pi_0(\mathbf{x})$ is a fixed reference pressure,
and neglecting terms that are nonlinear in $\pi_1$.
However, with this approach the contributions to the pressure from the fluid and from the magnetic field are treated on an equal footing,
whereas in reality we expect the magnetic pressure to be only a small perturbation to a non-magnetic reference state;
this is equivalent to the assumption that $v \ll c$.
We will therefore make a double approximation
in which we write $\pi = p_0(\mathbf{x}) + \pi_1$
and then neglect terms that are nonlinear in $\pi_1$ or $s_B$.
The Lagrangian density then becomes
\begin{align}
  \mathcal{L}
  &= \tfrac{1}{2}\rho|\mathbf{u}|^2 - \rho\Phi - \rho \tilde{H}(p_0,s,0) - \pi_1\rho \frac{\partial\tilde{H}}{\partial\pi}(p_0,s,0) - s_B\rho \frac{\partial\tilde{H}}{\partial s_B}(p_0,s,0) + p_0 + \pi_1\,.
\end{align}
Using the definitions given above, this can be rewritten in the form
\begin{align}
  \mathcal{L}
  &= \tfrac{1}{2}\rho|\mathbf{u}|^2 - \rho\Phi - \rho H(p_0,s) + p_0
  - \frac{\rho^\star}{\rho}\dfrac{|\mathbf{B}|^2}{2\mu_0} + \pi_1(1 - \rho/\rho^\star)\,,
  \label{eq:PI-MHD}
\end{align}
where
\begin{equation}
  \rho^\star(p_0,s) \equiv \left.1\middle/\frac{\partial H}{\partial p}(p_0,s)\right.
  \label{eq:PI_EoS}
\end{equation}
is the density given by the usual equation of state, but with $p$ replaced by $p_0$.
When we apply Hamilton's principle to this action,
the pressure perturbation $\pi_1$ serves as a Lagrange multiplier that enforces the approximate equation of state $\rho = \rho^\star(p_0,s)$.
The equation of motion is eventually found to be
\begin{equation}
  \rho\frac{\DD\mathbf{u}}{\DD t} = -\rho\nablab\Phi - \nablab p_0
  + \frac{1}{\mu_0}(\nablab\times\mathbf{B})\times\mathbf{B}
  - \nablab p_1 + \frac{p_1}{\rho c^2}\nablab p_0\,,
\end{equation}
where $p_1 \equiv \pi_1 - \dfrac{|\mathbf{B}|^2}{2\mu_0}$
and $\dfrac{1}{c^2} = \dfrac{\partial\rho^\star}{\partial p}(p_0,s)$.
Thus we finally arrive at the same set of pseudo-incompressible equations as in the non-magnetic case, except that the Lorentz force is now included in the momentum equation in its usual form,
and the induction equation also appears in its usual form
(as guaranteed by the definition of $\delta\mathbf{B}$ in equation~(\ref{eq:xi})).
From the action given by equation~(\ref{eq:PI-MHD}),
we deduce that this system conserves the same energy,
\begin{equation}
  \int\left[\tfrac{1}{2}\rho|\mathbf{u}|^2 + \rho\Phi + \rho U(\rho,s) + \frac{|\mathbf{B}|^2}{2\mu_0}\right]\,\dd^3\mathbf{x},
\end{equation}
as the fully compressible system.
The fact that the equations can be obtained from an action
also implies that the linearised equations are self-adjoint (in the absence of rotation).

In order to include diabatic processes in the model,
it is necessary to relate the temperature to the other thermodynamic variables.
However, since the density is now given by the approximate equation of state~(\ref{eq:PI_EoS}),
the correct definition of temperature is not obvious.
In fact, there is a unique definition that preserves the laws of thermodynamics \citep{KleinPauluis12}:
\begin{equation}
  T = T^\star(p_0,s) + p_1 \, \frac{\partial T^\star}{\partial p}(p_0,s)\,,
\end{equation}
where $T^\star(p,s) \equiv \dfrac{\partial H}{\partial s}(p,s)$ is the usual equation of state.
This is the definition that we have used in our results; in its linearised form, it is identical to the fully compressible relation~(\ref{eq:temp}).

\bibliography{Bibliography}

\end{document}